\documentclass[twoside,a4paper,11pt]{sea10}
\usepackage{graphicx}
\usepackage{hyperref}
\begin{document}
\pagenumbering{arabic}
\pagestyle{myheadings}
\thispagestyle{empty}
{\flushleft\includegraphics[width=\textwidth,bb=58 650 590 680]{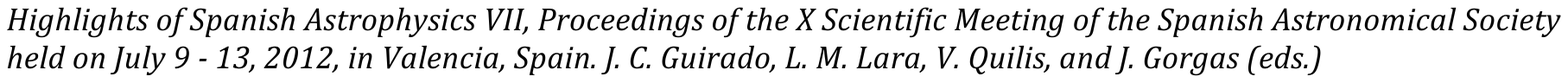}}
\vspace*{0.2cm}
\begin{flushleft}
{\bf {\LARGE
%
Witnessing galaxy clusters: from maturity to childhood 
%
}\\
\vspace*{1cm}
%
B. Ascaso$^{1}$
%
}\\
\vspace*{0.5cm}
%
$^{1}$
Instituto de Astrof\'isica de Andaluc\'ia (IAA-CSIC), Glorieta de la Astronom\'ia s/n, E-18008 Granada, Spain
%
\end{flushleft}
%
\markboth{
Witnessing galaxy clusters: from maturity to childhood 
}{ 
%
B. Ascaso
%
}
\thispagestyle{empty}
\vspace*{0.4cm}
\begin{minipage}[l]{0.09\textwidth}
\ 
\end{minipage}
\begin{minipage}[r]{0.9\textwidth}
\vspace{1cm}
\section*{Abstract}{\small
Galaxy clusters are the largest structures in Universe. They are very important as both cosmological probes and astrophysical laboratories.  Several methods have been developed to detect galaxy clusters with different techniques (optical, X-rays, Weak Lensing and Sunyaev-Zeldovich effect) providing cluster samples with a well-characterized purity and completeness rates up to moderate redshift (z$<$1.2). These samples allow us to study the systematic of different methods and to obtain reliable mass estimations. On the contrary, high redshift clusters only started to be explored very recently with the advent of deep IR and X-ray data surveys, providing the first proto-clusters (z$>$1.5-2) ever detected. In this talk, I introduce these techniques and review some of the cluster samples obtained including particular striking cases. I discuss their relevance in terms of cosmological and galaxy evolution constraints and finally, I briefly refer to the cluster science predictions for the next generation surveys.
%
\normalsize}
\end{minipage}
%
%
%

\section{What are galaxy clusters?}

Galaxy clusters are the largest structures gravitationally bounded in the Universe. They contain tens to thousands of galaxies, hot gas or plasma and a large percentage of dark matter, detected through indirect ways. Understanding the way that these objects were formed and are evolving is very important since this is directly related to the formation and evolution of the Universe.

Assuming a cosmology scenario where density perturbations were generated in the early universe and amplified by gravity \cite{press74,peebles93,peacock99}, we expect to observe forming galaxy clusters at high redshift and therefore, we should be able to trace their evolution from this redshift to the present time. Hence, the discoveries of such clusters and their descendants are crucial to restrict the formation and evolution of such structures. 

On the other hand, many numerical simulations have intended to model the processes of formation and evolution of galaxy clusters through numerical and N-body simulations \cite{diemand05,springel05,planelles09,stanek09,sembolini12} for different components of the cluster. The comparison of these simulations with observations provides constraints on the main theories of formation and evolution of Universe.

\section{How can we find them?}

Galaxy clusters occupy very massive dark matter halos. The most massive ones are easy to identify up to moderate redshift since they contain large numbers of tightly clustered galaxies \cite{depropris02,ascaso08,ascaso09}, strong X-ray emission signatures \cite{ebeling96,rosati02,bohringer04,bohringer07}, relatively strong features in the gravitational lensing shear field \cite{wittman01,wittman03,zitrin09} and potential Sunyaev-ZelÕdovich signatures \cite{ascaso07,menanteau09,menanteau10}.

\subsection{Light tracers}

Cluster detection methods based on optical data have provided a large dataset of clusters. They take advantage of the larger number of bands, better photometric redshift quality or improvement in depth of the new surveys. We summarize below the main optical cluster detection methods in three main groups based on the use of different optical characteristics.

Many optical selection techniques have been published: based on spatial distribution alone (Voronoi tessellation; \cite{ramella01,lopes04,soares-santos11}, on luminosity and density profiles (matched and adaptive filters, \cite{postman96,postman02,kepner99}, on constituent galaxy colors and the magnitude of the BCG (ie. red sequence: \cite{gladders00,gilbank11}; MaxBCG: \cite{koester07}; GMBCG: \cite{hao10}), and/or on photometric redshifts (eg. \cite{olsen07,menanteau09,wen12}). 

We created an innovative technique that we developed to detect galaxy clusters in the optical: the Bayesian Cluster Finder (BCF; \cite{ascaso12}). The BCF computes the probability of a cluster having a given luminosity, density, and photometric redshift distribution profile. Furthermore, this method allows one to find galaxy clusters without a predefined CMR or a given BCG, while using this information for the enhancement of this probability as a prior term. 
                             
The BCF can be also extended to higher redshift ranges (z$>$1) by using deep Infrared data. Similar works \cite{eisenhardt08,wilson09}, have also detected galaxy clusters in the high redshift regime with different techniques. The SpARCS Survey \cite{wilson09}, which aimed to detect galaxy clusters based on the presence of the red sequence, has just published a few detections. The main concern about this method is that the detections are likely to be biased towards the most virialized systems containing a well-formed red sequence at any redshift. On the contrary, \cite{eisenhardt08} used a wavelet analysis independent of the colors of the galaxies to detect galaxy clusters. They recovered a mean of $\sim$14 detections per square degree at z$>$1 and estimated a maximum of 10\% of false detections. In addition, new ways of detecting methods at higher redshift have been developed for detecting galaxy clusters around radio galaxies at z$\sim$ 1.5-2 \cite{galametz09,chiaberge10}. 

\subsection{Gas tracers}

\subsubsection{X-rays}

A number of methods to detect clusters have been developed based on cluster X-ray emission \cite{rosati02}. One of the first methods designed to analyze EINSTEIN and ROSAT data was the sliding cell method. This method selects sources based on a signal-to-noise threshold which is estimated from positive fluctuations that deviate significantly from a background map. This map is estimated from a fixed size cell sliding across the image. Even if this method is performing well in point sources, it is not optimized for extended sources producing low completeness rates for deeper X-ray data.

Different methods were developed to take into account the extended emission nature of the data. One of the most popular was the wavelet techniques \cite{rosati95,lazzati99,romer00}. This method consists of enhancing the source signal versus the background by decomposing the signal via the wavelet transform.  It reaches good completeness and purity rates and provides measurements of the source parameters. Other well-known methods are the Voronoi Tessellation and Percolation \cite{ebeling93,scharf97}, the growth curve analysis (GCA, \cite{bohringer00}) or the identification of X-ray clusters from two-band imaging with  final spectroscopic confirmation \cite{fassbender08} among others. Even if all of them deal with the extent of the sources, their completeness and purity rates change accordingly with their availability to confirm spectroscopically the detections. 

While the X-rays have been successful at detecting high-z clusters as in the XMM Newton Distant Cluster Project (XMM-NDCP, \cite{fassbender08}), the structures found are usually very massive.

\subsubsection{Sunyaev-Zel'dovich effect}

The Sunyaev-Zel'dovich (SZ) effect is also an excellent tracer of the gas of the cluster. This effect is produced when photons that come from the cosmic microwave radiation and pass through a cloud of electrons contained in the cluster gas, result into a shift of the black body energy spectrum \cite{sunyaev72}. This shift produces a decrease at lower energies and an increase at higher energies.

The SZ effect, even if it is independent of the redshift of the cluster, it does depend on the mass of the cluster \cite{ascaso07}. Hence, since clusters are predicted to be growing in the standard $\Lambda$CDM scenario, only very rare structures at high-z are expected to be detected with this technique.  Semi-analytic, hydrodynamic and N-body simulations \cite{white03,schulz03,vale06} have been used to test the reliability of the detection methods, such as matched filter, Fourier and wavelet based techniques. The results show that the completeness and efficiency depend strongly on the frequency coverage, the number of frequencies observed, the point source confusion and the candidate mass.

Many experiments have been designed to study galaxy clusters with the SZ effect. Three of the best known experiments that have already produced cluster catalogs are the Atacama Cosmology Telescope (ACT, \cite{menanteau10}), the South Pole Telescope (SPT, \cite{staniszewski09}) and PLANCK \cite{planck11}. Their study of their properties and systematics is just starting (e.g. \cite{menanteau10b,marrone12}). 

\subsection{Dark matter tracers}

\begin{figure}
\center
\includegraphics[width=8.3cm,angle=0,clip=true]{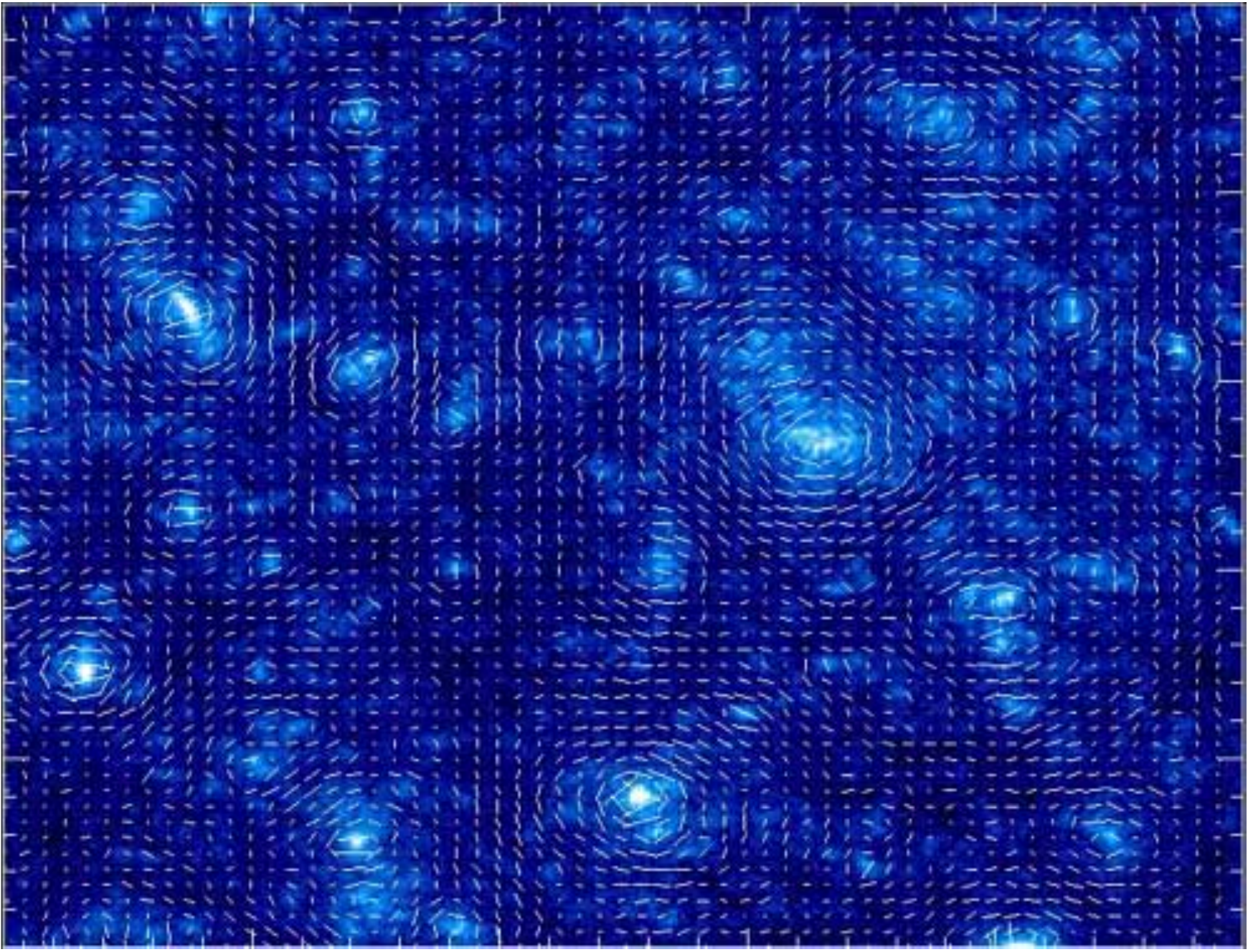} 
\caption{\label{fig:WLsimu} Simulation of cosmic shear in a Cold Dark Matter (CDM) cosmological model extracted from \cite{jain00} (more information can be found there). The size of this figure is 0.5x0.5 square degrees and the lighter regions represent over-densities corresponding to galaxy groups and clusters. The segments represent the amplitude and direction of the shear produced by the structures.}
\end{figure}

A different way to detect galaxy clusters is to investigate the presence of dark matter by analyzing the cosmic shear produced by the weak lensing (WL) effect \cite{tyson90,wittman01,wittman03} in a survey. Figure \ref{fig:WLsimu} shows a simulation of cosmic shear in a CDM cosmological model \cite{jain00}. We see how the most massive structures create a shear pattern on the galaxies around, which is produced by the massive dark matter structure existing at this location.

\begin{figure}
\center
\includegraphics[width=8.3cm,angle=0,clip=true]{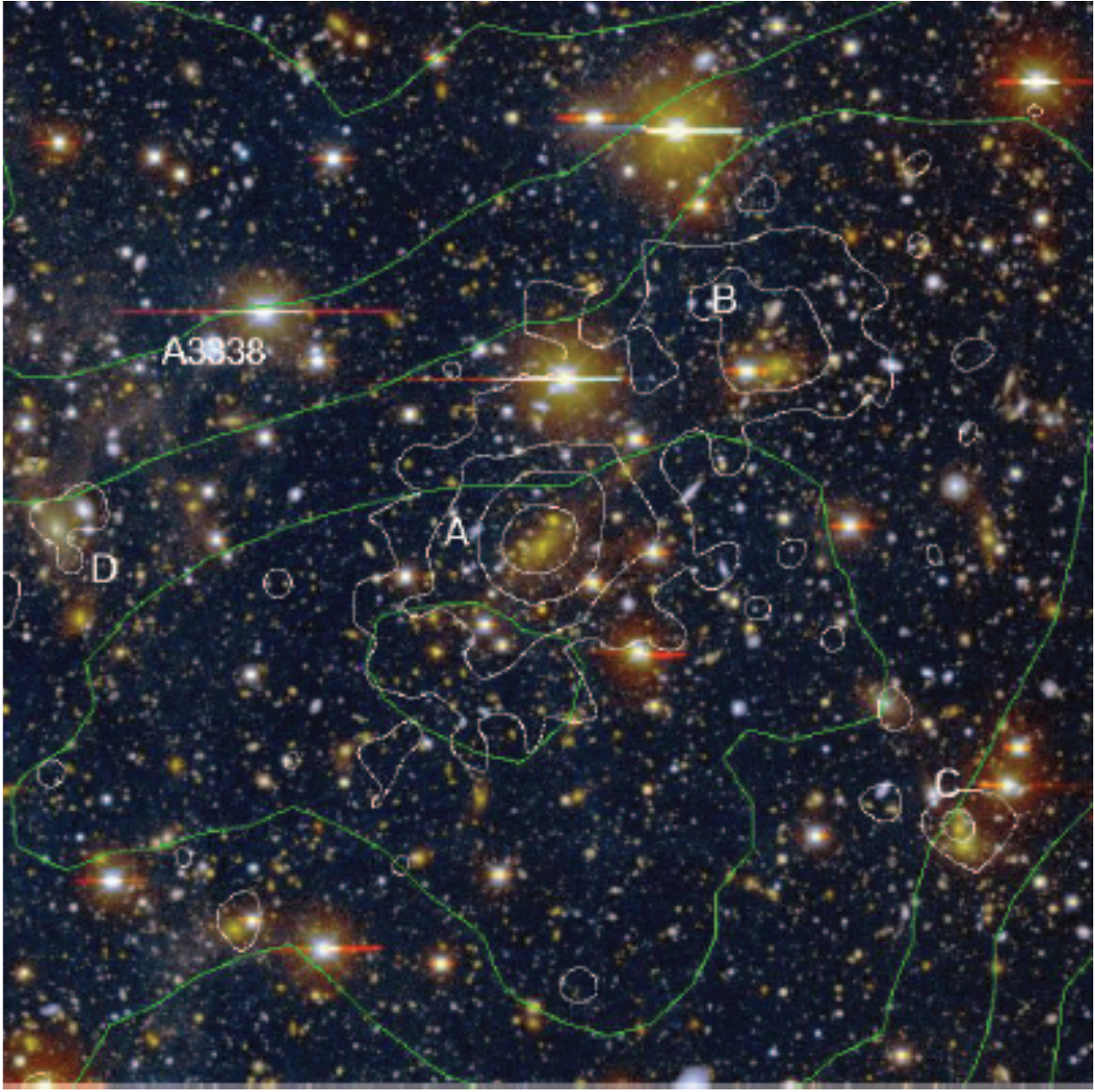} 
\caption{\label{fig:WLclus} One of the clusters detected and confirmed in the DLS: DLSCL J0522.24820. The color image shows the cluster in the optical, while the white contours refer to the the X-ray detections and green contours refer to the WL signal. More information in \cite{wittman06}.}
\end{figure}

Many present surveys are devoted to the study of clusters detected via WL convergence maps (e.g. the Deep Lens survey, \cite{wittman02,wittman06}, the CFHTLS, \cite{gavazzi07,shan12}). In Figure \ref{fig:WLclus}, we show the optical counterpart of a galaxy cluster detected through weak lensing in the DLS \cite{wittman06}.

In general, different techniques and tracers provide different set of detections even if they agree at the high mass end (Ascaso et al. 2012b, in prep). The understanding of these systematics is crucial to understand the main nature of galaxy clusters.

\section{Which kind of clusters do we find?}

Galaxy clusters are both cosmological probes, very useful to constrain cosmological parameters and astrophysical laboratories, ideal to study galaxy evolution. In this section, we review some of the main highlights of galaxy clusters at different redshift ranges separately.  We consider 'adult'  ($z<1$), 'teenagers' ($1<z<2.5$) and 'newborn' ($z>2.5$) clusters.

\subsection{$z<1$ clusters}

A large number of cluster catalogues are available up to redshift 1 from different surveys and techniques as the ones specified above. In the optical, even if wide surveys are always desirable, a compromise needs to be achieved between width, depth and photometric redshift resolution (number of bands) due to telescope time-consuming reasons. In general, spectroscopic surveys are able to detect galaxy clusters and groups down to the only few galaxies. However, they are usually shallower than photometric surveys and therefore, their redshift completeness is moderate. Reversely, just photometric surveys arrive up to z$\sim$1 or even higher, but their low photometric redshift resolution, if any, results into a high mass sample. Intermediate approaches like photometric surveys with large number of narrow bands such as the present ALHAMBRA survey \cite{moles08} or the next generation J-PAS survey \cite{benitez09} produce relatively deep data (high redshift clusters) and samples well the cluster and group mass function (Ascaso et al. 2012c, in prep).

Other specific cluster surveys provide a careful look into different properties of a particular set of clusters \cite{white05,postman12}. In particular,  CLASH \cite{postman12} consists of 25 clusters imaged with the ACS/WFC3 with 16 bands from the UV to the IR.

\begin{figure}
\center
\includegraphics[width=8.3cm,angle=0,clip=true]{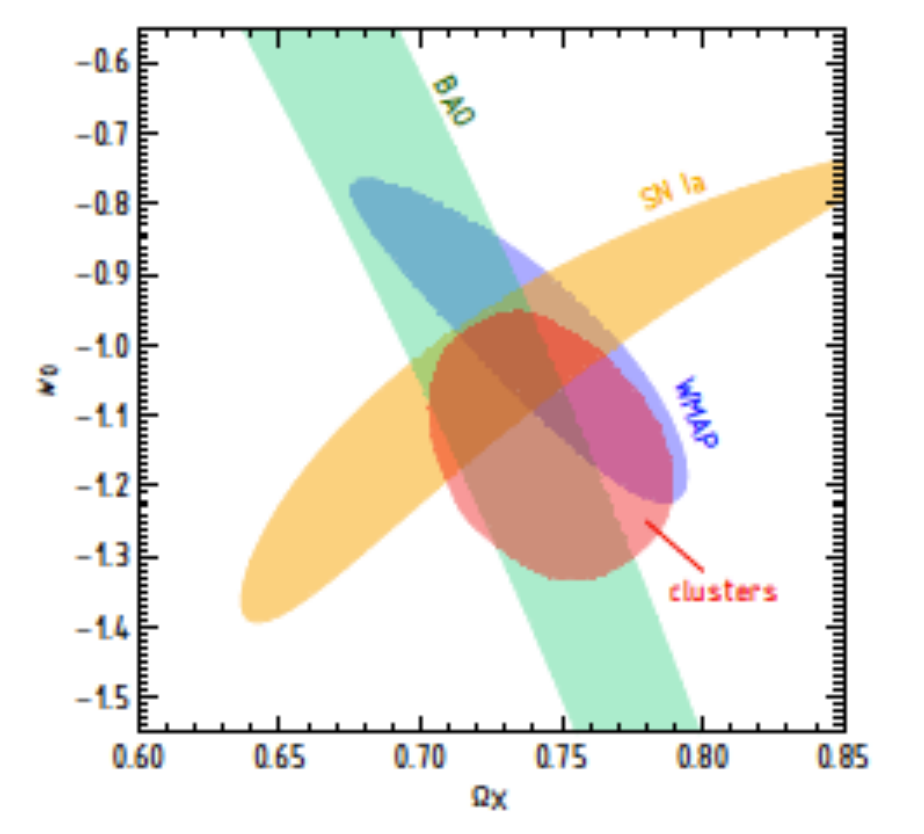} 
\caption{\label{fig:cosmology} Figure extracted from \cite{vikhlinin09} (more information can be found there). Comparison of the dark energy constraints from X-ray clusters, SNe, BAOs, and WMAP).}
\end{figure}

Cosmologically, precise observations of large numbers of galaxy clusters are powerful instruments for our understanding of cosmology and structure formation. The abundance of galaxy clusters is extremely sensitive to the amplitude of matter fluctuations and variations in cosmology according to theoretical predictions (e.g \cite{press74,stanek09}). Several works in the X-rays \cite{vikhlinin09}, SZ \cite{carlstrom02}, WL \cite{johnston07} and optical \cite{wen10,rozo10} have used clusters to constrain cosmological parameters. The constraints that the evolution of clusters provides, together with constraints coming from the Cosmic Microwave Background (CMB), Supernovae (SNe) or Baryonic Acoustic Oscillations (BAOs) can help to reduce significantly the error of the estimation for cosmological parameters, in particular dark energy parameters (see Fig  \ref{fig:cosmology}). 

On the other hand, a number of studies of individual massive clusters in this range of redshift have measured the properties of the galactic population in clusters. Due to to time limitations, I only mention here some results, but many other references can be found in the literature. For instance, many works have well characterized the slope of the color-magnitude relation at least up to redshift 1.6 \cite{mei06,mei09,ascaso08,papovich10}, proved the blue fraction evolution even with a wide dispersion \cite{margoniner01,ascaso08}, studied structural parameter evolution of galaxies in clusters \cite{ascaso09,ascaso11} and well measured and fit the luminosity function down to the faint end for different surveys \cite{blanton03,harsono09}. 

\subsection{$1<z<2.5$ clusters}

As mentioned before, two main deep IR surveys have systematically searched for galaxy clusters \cite{eisenhardt08,wilson08,wen11} at z$>$1. In addition, the XMM-NDCP \cite{fassbender08} provided a X-ray selected sample of galaxy clusters.

The cosmological implications of these high-z cluster catalogs are numerous. For instance, \cite{hoyle11} analyzed a sample of 14 spectroscopically confirmed clusters at z$>$1 with mass measurements. They explored whether the abundance of these observed clusters was compatible with the standard $\Lambda$CDM scenario. They found that there was a discrepancy which could be attenuated by considering non-Gaussian primordial perturbations of local type, $f_{NL}$ at a 3$\sigma$ level. Otherwise, if $f_{NL}$ did not exist, the measured masses should be systematically lower or $\sigma_8$ higher that the present predictions. Hence, the need of revisiting the $\Lambda$CDM paradigm or the uncertainties in the observations is already arising with z$>$1 cluster sets.

\begin{figure}
\center
\includegraphics[scale=0.5]{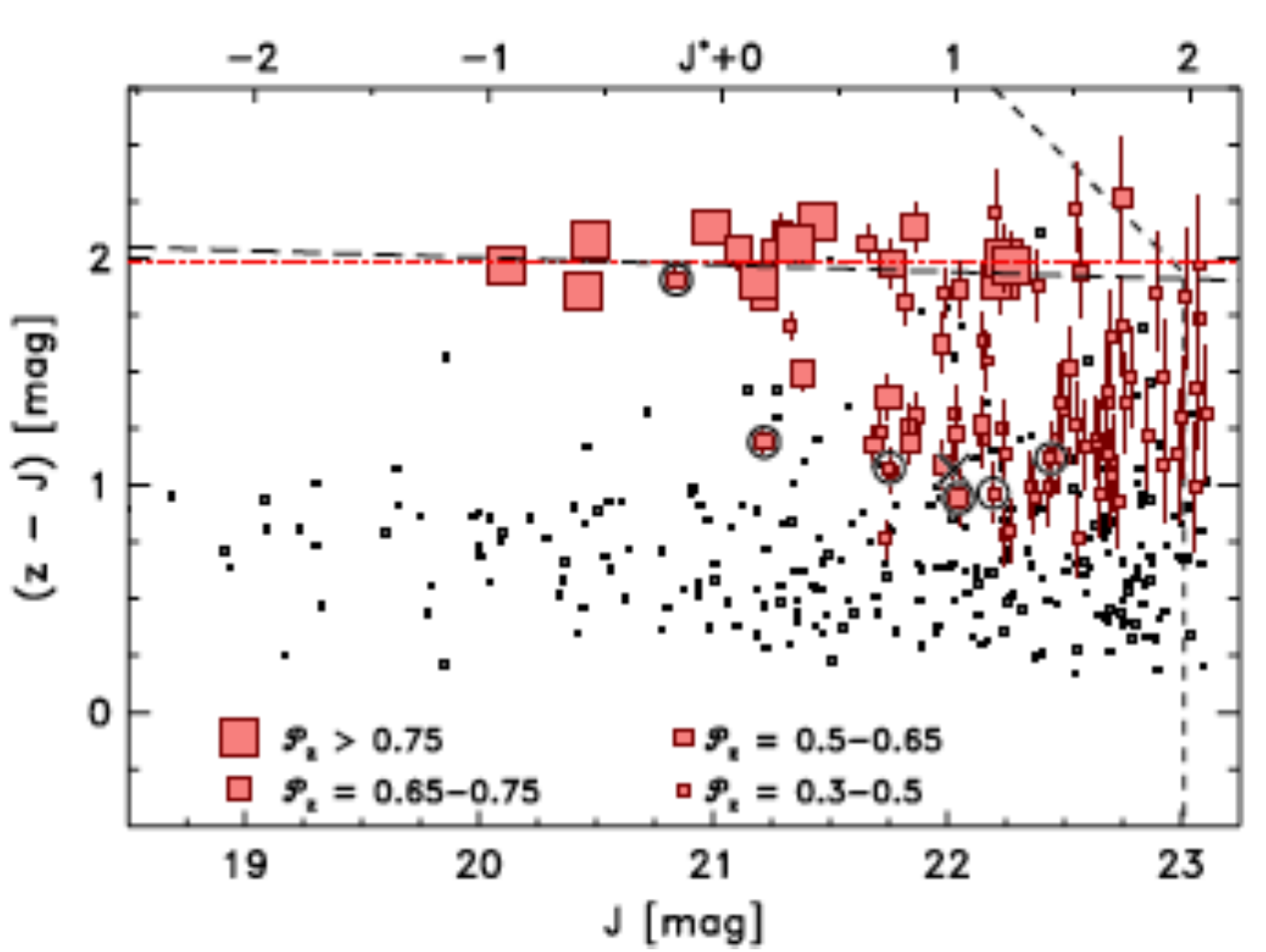} ~
\includegraphics[scale=0.5]{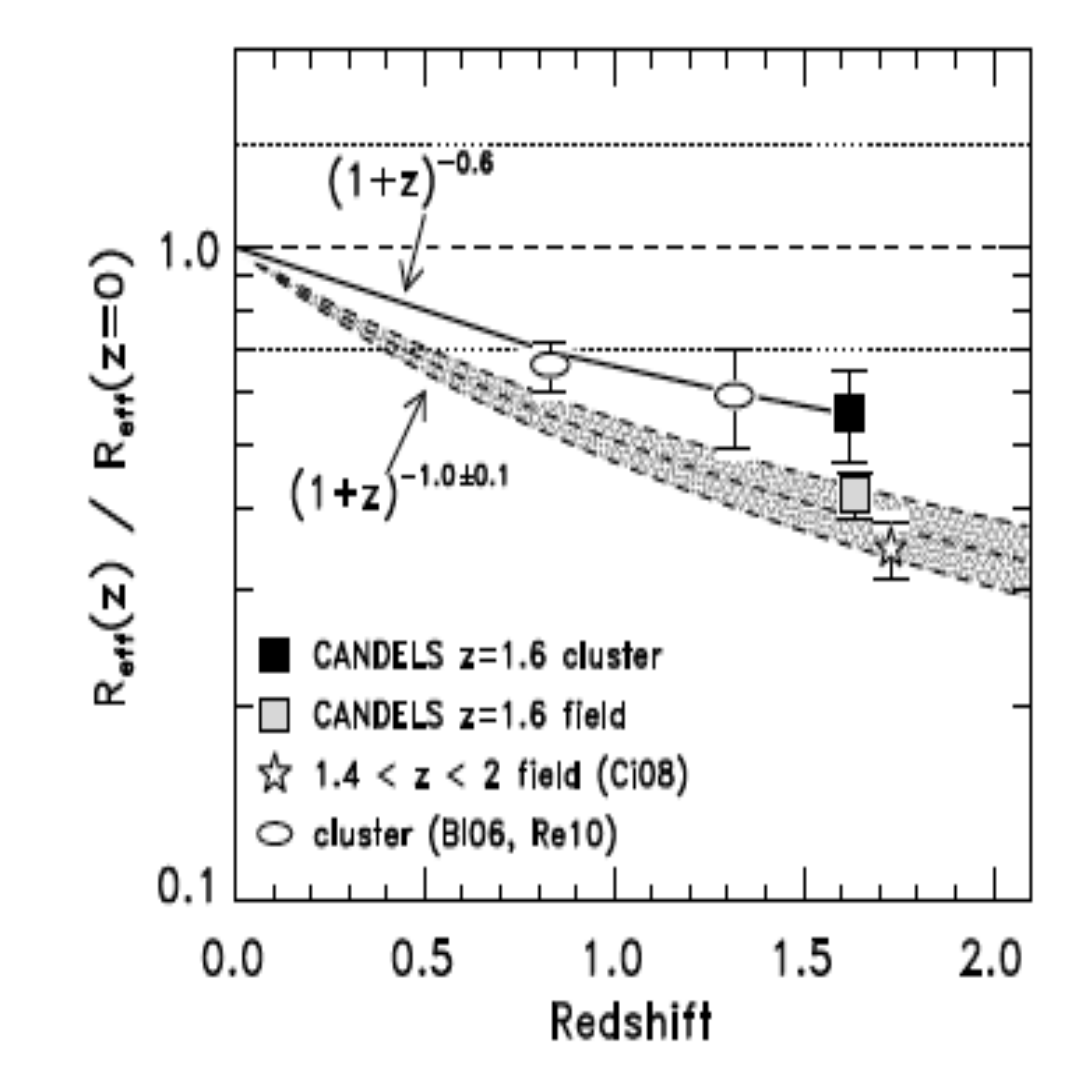} ~
\caption{\label{fig:papovich} (Left) Figure extracted from \cite{papovich10} (a detailed description can be found there). Color-magnitude diagram of all galaxies within 1 Mpc. The long-dashed line is the fit to the well-defined red sequence. (Right) Figure extracted from \cite{papovich12} (a detailed description can be found there). Comparison of the evolution of the effective radii for massive cluster galaxies. The thick, solid line shows the size evolution measured for the cluster galaxies whether the shaded curve indicates the size evolution for a mix of field and cluster early-type galaxies from van der Wel et al. (2008). It appears that cluster galaxies have had a milder size evolution since z$<$1.6.} 
\end{figure}

As far as galaxy evolution at this range of redshift is concerned, few clusters have been studied in detail. One of the few examples that has been extensively studied and spectroscopically confirmed is ClG J0218.3-0510, at z=1.62, detected in the SWIRE IR Survey. \cite{papovich10} showed that this cluster has a prominent red sequence, dominated by a population of red galaxies (see Fig \ref{fig:papovich},(left)). In addition, \cite{papovich12} analyzed the structural and morphological properties of the galactic population of this cluster, finding a color-morphology relation very similar to what it is found at lower redshift. Moreover, they measure the size evolution of the massive cluster galaxies from z=1.62 to present (See Fig \ref{fig:papovich},(right)), which is directly related to the constraint of possible evolutionary scenarios. Another example is JKCS 041at z$\sim$ 2.2, detected through the red sequence technique and confirmed by X-ray data \cite{andreon09}. This cluster has a very tight red sequence \cite{andreon11} as well, and no evidence of the BO effect has been found for the most massive galaxies \cite{raichoor12}. A detailed analysis of this cluster and its properties can be found in \cite{andreon09} and references herein. A wider variety of clusters at this range of redshift is necessary to be able to constrain evolutionary processes.

\subsection{$z>2.5$ clusters}

\begin{figure}
\center
\includegraphics[width=8.3cm,angle=0,clip=true]{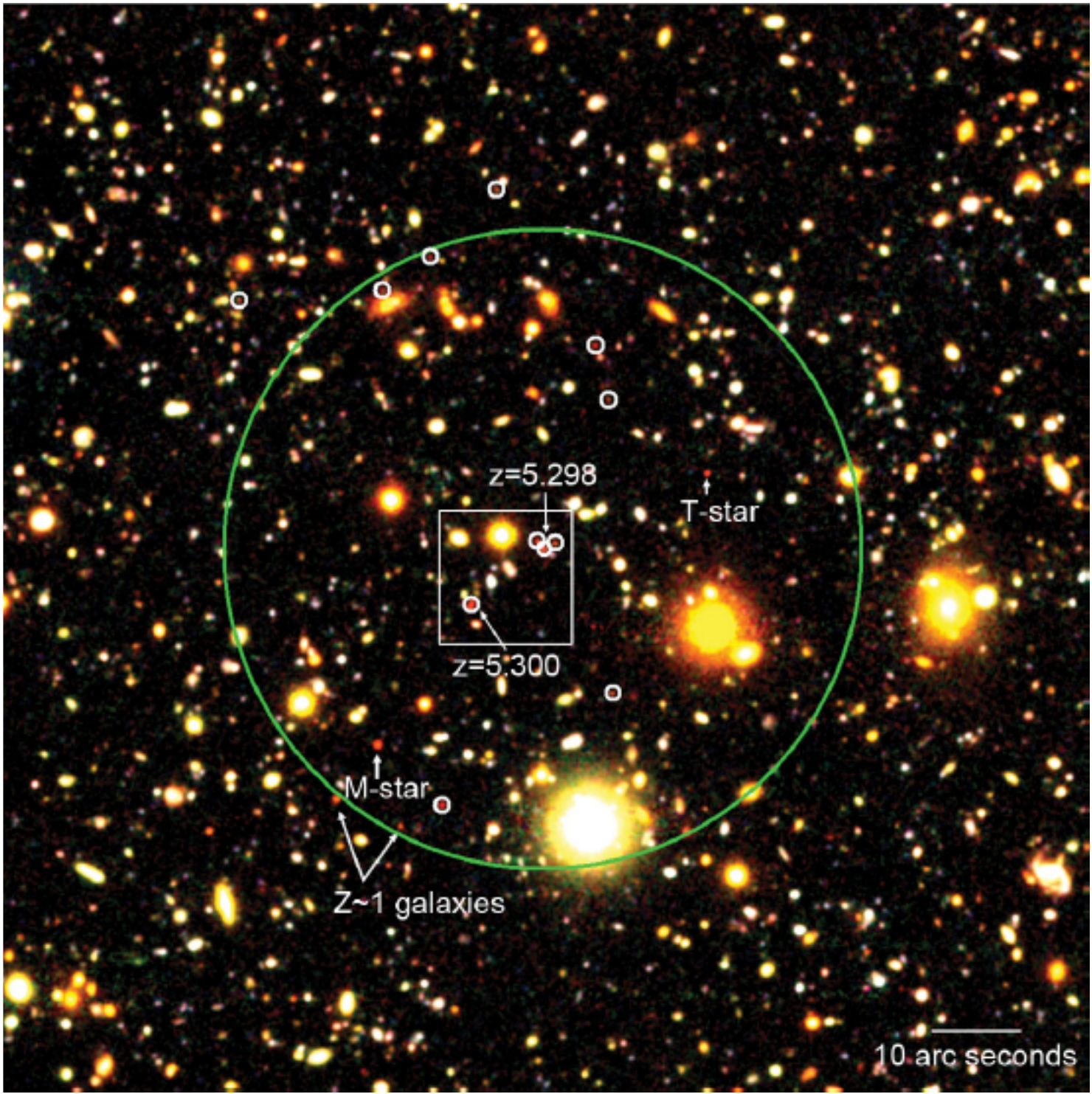} 
\caption{\label{fig:protocluster}  Figure extracted from \cite{capak11} (a more detailed description can be found there). Proto-cluster at z$\sim$5.3. The green circle refers to the comoving radius of 2 Mpc around the starburst COSMOS AzTEC-3. The z $\sim$5.3 candidates are marked in white.}
\end{figure}

According to CDM cosmological simulations, massive galaxies would merge hierarchically to form a cluster \cite{springel05}. The early stages of this process is usually known as proto-clusters.  At z$>$2.5, no systematic search for such clusters exists.  In general, these clusters are usually detected as overdensities of galaxies around really massive galaxies such as luminous quasars and starburst and confirmed spectroscopically later. Figure \ref{fig:protocluster}, shows one of the most distant proto-cluster up to date \cite{capak11} at  z$\sim$ 5.3, just $\sim$1 Gyr after the Big Bang. It contains a luminous quasar and a system rich in molecular gas. The lower mass limit measured for this cluster is 4$\times 10^{11} M_{\odot}$, which is consistent with the expectations from the CDM simulations.

Even if the number of proto-clusters detections do not exceed a decen \cite{daddi09,kuiper11,carilli11,walter12,toshikawa12}, it seems that their general properties such as mass or luminosity agree with the values expected from hierarchical simulations. The properties of their galactic population have barely been studied since it is very complicated to confirm the membership of such clusters. Larger proto-cluster samples are needed in order to extract reliable results.

\section{Which is the role of galaxy clusters in the upcoming surveys?}

In the coming years, a number of large surveys will provide enormous amount of data. Even if each survey follows a different strategy according to their main scientific objectives, almost all of them are collecting large areas of the sky. While few of these surveys will be spectroscopic (BOSS, BigBOSS), some others will use broad band imaging (DES, Pan-STARSS, LSST, Euclid) and others, such as J-PAS will use narrow band imaging with 56 filters which will provide 0.3\% photo-z accuracy.

As discussed before, spectroscopy or a good photometric redshift precision will lead to detect and sample the low-mass part of the cluster and group distribution. Additionally, deep photometric infrared data (or even longer wavelengths) is needed to arrive up to the highest redshift possible and obtain larger samples of clusters at z$>$1. Starting in 2014, J-PAS \cite{benitez09} seems one of the most feasible survey to get both an excellent photometric resolution and depth, even in the IR. Preliminary simulations (Ascaso et al. 2012c) predict that J-PAS will be able to detect $\sim 1.5\times 10^6$ clusters and groups, up to z$<$1.5, and down to the level of few galaxies. Additional searches of very high redshift radio sources will complement the high redshift end. 

This sample and similar ones that will be obtained in the future years will make a substantial change in the cluster paradigm as we know it today. The cosmological models will suffer a exhaustive revision by the new sets of observational data and excellent techniques which will presumably solve and constrain important enigmas such as the nature of the dark energy, the understanding of the mechanisms behind the galaxy evolution in clusters and many others.

%
%
\small  
%
\section*{Acknowledgments}   
%

I would like to thank Javier Gorgas and the SOC of the meeting for inviting me to give this talk and to the LOC for organizing such a wonderful meeting.  I am also grateful to my present scientific group for support and inspiration: Txitxo Ben\'itez, Alberto Molino, Yoli Jim\'enez-Teja and William Schoenell. Thanks to Stefano Andreon and JinLin Han for suggestions. Lastly, I would like to thank my collaborators from the DLS, the ALHAMBRA, the CLASH and the J-PAS teams since most of the content of this talk comes from learning and working with them. 
%

%

\begin{thebibliography}{}
\small
%


\bibitem{andreon09} Andreon, S. et al.\ 2009, A\&A, 507, 147 
\bibitem{andreon11} Andreon, S., \& Huertas-Company, M.\ 2011, A\&A, 526, A11 
\bibitem{ascaso07} Ascaso, B., \& Moles, M.\ 2007, ApJL, 660, L89
\bibitem{ascaso08} Ascaso, B. et al.\ 2008, A\&A, 487, 453
\bibitem{ascaso09} Ascaso, B. et al. \ 2009, A\&A, 506, 1071
\bibitem{ascaso11} Ascaso, B. et al.\ 2011, ApJ, 726, 69
\bibitem{ascaso12} Ascaso, B., Wittman, D., \& Ben{\'{\i}}tez, N.\ 2012, MNRAS, 420, 1167 
\bibitem{benitez09} Ben{\'{\i}}tez, N., et al.\ 2009, ApJ, 691, 241
\bibitem{blanton03} Blanton, M.~R. et al.\ 2003, ApJ, 592, 819
\bibitem{bohringer00} B{\"o}hringer, H. et al.\ 2000, ApJS, 129, 435 
\bibitem{bohringer04} B{\"o}hringer, H. et al.\ 2004, A\&A, 425, 367
\bibitem{bohringer07} B{\"o}hringer, H. et al.\ 2007, A\&A, 469, 363
\bibitem{capak11} Capak, P.~L., et al.\ 2011, Nature, 470, 233 
\bibitem{carilli11} Carilli, C.~L. et al.\ 2011, ApJL, 739, L33
\bibitem{carlstrom02} Carlstrom, J.~E., Holder, G.~P., \& Reese, E.~D.\ 2002, ARA\&A, 40, 643
\bibitem{chiaberge10} Chiaberge, M. et al., ApJL, 710, L107 
\bibitem{daddi09} Daddi, E. et al.\ 2009, ApJ, 694, 1517 
\bibitem{depropris02} De Propris, R. et al.\ 2002, MNRAS, 329, 87
\bibitem{diemand05} Diemand, J. et al.\ 2005, MNRAS, 364, 665 
\bibitem{ebeling93} Ebeling, H., \& Wiedenmann, G.\ 1993, PhRvE, 47, 704
\bibitem{ebeling96} Ebeling, H. et al.\ 1996, MNRAS, 281, 799
\bibitem{eisenhardt08} Eisenhardt, P.~R.~M., et al.\ 2008, ApJ, 684, 905 
\bibitem{fassbender08} Fassbender, R.\ 2008, arXiv:0806.0861 
\bibitem{galametz09} Galametz, A., et al.\ 2009, A\&A, 507, 131 
\bibitem{gavazzi07} Gavazzi, R., \& Soucail, G.\ 2007, A\&A, 462, 459
\bibitem{gilbank11} Gilbank, D.~G. et al.\ 2011, AJ, 141, 94
\bibitem{gladders00} Gladders, M.~D., \& Yee, H.~K.~C.\ 2000, AJ, 120, 2148 
\bibitem{hao10} Hao, J. et al.\ 2010, ApJS, 191, 254
\bibitem{harsono09} Harsono, D., \& De Propris, R.\ 2009, AJ, 137, 3091
\bibitem{hoyle11} Hoyle, B., Jimenez, R., \& Verde, L.\ 2011, PhRvD, 83, 103502 
\bibitem{jain00} Jain, B., Seljak, U., \& White, S.\ 2000, ApJ, 530, 547
\bibitem{johnston07} Johnston, D.~E. et al.\ 2007, arXiv:0709.1159 
\bibitem{kepner99} Kepner, J. et al.\ 1999, ApJ, 517, 78 
\bibitem{koester07} Koester, B.~P., et al.\ 2007, ApJ, 660, 221 
\bibitem{kuiper11} Kuiper, E. et al.\ 2011, MNRAS, 417, 1088
\bibitem{lazzati99} Lazzati, D. et al.\ 1999, ApJ, 524, 414 
\bibitem{lopes04} Lopes, P.~A.~A. et al.\ 2004, AJ, 128, 1017 
\bibitem{margoniner01} Margoniner, V.~E., de Carvalho, R.~R., Gal, R.~R., \& Djorgovski, S.~G.\ 2001, ApJL, 548, L143
\bibitem{mei06} Mei, S., et al.\ 2006, ApJ, 644, 759 
\bibitem{mei09} Mei, S., et al.\ 2009, ApJ, 690, 42 
\bibitem{marrone12} Marrone, D.~P. et al.\ 2012, ApJ, 754, 119 
\bibitem{menanteau09} Menanteau, F., \& Hughes, J.~P.\ 2009, ApJL, 694, L136
\bibitem{menanteau10} Menanteau, F. et al.\ 2010, ApJ, 723, 1523 
\bibitem{menanteau10b} Menanteau, F. et al.\ 2010, ApJS, 191, 340 
\bibitem{moles08} Moles, M. et al.\ 2008, AJ, 136, 1325
\bibitem{olsen07} Olsen, L.~F., et al.\ 2007, A\&A, 461, 81
\bibitem{papovich10} Papovich, C. et al.\ 2010, ApJ, 716, 1503
\bibitem{papovich12} Papovich, C. et al.\ 2012, ApJ, 750, 93 
\bibitem{peacock99} Peacock J.A.. 1999. Cosmological Physics. Cambridge, UK: Cambridge Univ. Press
\bibitem{peebles93} Peebles, P.J.E.,\ 1993, Physical Cosmology, Princeton, NJ: Princeton Univ. Press
\bibitem{planck11} Planck Collaboration et al.\ 2011, A\&A, 536, A8
\bibitem{planelles09} Planelles, S., \& Quilis, V.\ 2009, MNRAS, 399, 410 
\bibitem{postman96} Postman, M. et al.\ 1996, AJ, 111, 615 
\bibitem{postman02} Postman, M. et al.\ 2002, ApJ, 579, 93
\bibitem{postman12} Postman, M. et al.\ 2012, ApJS, 199, 25 
\bibitem{press74} Press, W.~H., \& Schechter, P.\ 1974, ApJ, 187, 425 
\bibitem{raichoor12} Raichoor, A., \& Andreon, S.\ 2012, A\&A, 537, A88
\bibitem{ramella01} Ramella, M. et al. \ 2001, A\&A, 368, 776
\bibitem{romer00} Romer, A.~K. et al.\ 2000, ApJS, 126, 209 
\bibitem{rosati95} Rosati, P. et al. \ 1995, ApJL, 445, L11 
\bibitem{rosati02} Rosati, P., Borgani, S., \& Norman, C.\ 2002, ARA\&A, 40, 539 
\bibitem{rozo10} Rozo, E. et al.\ 2010, ApJ, 708, 645
\bibitem{scharf97} Scharf, C.~A. et al.\ 1997, ApJ, 477, 79 
\bibitem{schulz03} Schulz, A.~E., \& White, M.\ 2003, ApJ, 586, 723
\bibitem{shan12} Shan, H. et al.\ 2012, ApJ, 748, 56 
\bibitem{sembolini12} Sembolini, F. et al.\ 2012, arXiv:1207.4438 
\bibitem{soares-santos11} Soares-Santos, M. et al.\ 2011, ApJ, 727, 45 
\bibitem{springel05} Springel, V. et al.\ 2005, Nature, 435, 629
\bibitem{stanek09} Stanek, R., Rudd, D., \& Evrard, A.~E.\ 2009, MNRAS, 394, L11
\bibitem{staniszewski09} Staniszewski, Z. et al.\ 2009, ApJ, 701, 32 
\bibitem{sunyaev72} Sunyaev, R.~A., \& Zeldovich, Y.~B.\ 1972, Comments on Astrophysics and Space Physics, 4, 173 
\bibitem{toshikawa12} Toshikawa, J. et al.\ 2012, ApJ, 750, 137 
\bibitem{tyson90} Tyson, J.~A., Wenk, R.~A., \& Valdes, F.\ 1990, ApJL, 349, L1
\bibitem{vale06} Vale, C., \& White, M.\ 2006, NewA, 11, 207 
\bibitem{vikhlinin09} Vikhlinin, A. et al.\ 2009, ApJ, 692, 1060
\bibitem{walter12} Walter, F. et al.\ 2012, Nature, 486, 233 
\bibitem{wen10} Wen, Z.~L., Han, J.~L., \& Liu, F.~S.\ 2010, MNRAS, 407, 533 
\bibitem{wen11} Wen, Z.~L., \& Han, J.~L.\ 2011, ApJ, 734, 68
\bibitem{wen12} Wen, Z.~L., Han, J.~L., \& Liu, F.~S.\ 2012, ApJS, 199, 34 
\bibitem{white03} White, M.\ 2003, ApJ, 597, 650 
\bibitem{white05} White, S.~D.~M. et al.\ 2005, A\&A, 444, 365 
\bibitem{wilson08} Wilson, G., et al.\ 2008, Infrared Diagnostics of Galaxy Evolution, 381, 210 
\bibitem{wilson09} Wilson, G., et al.\ 2009, ApJ, 698, 1943 
\bibitem{wittman01} Wittman, D. et al.\ 2001, ApJL, 557, L89 
\bibitem{wittman02} Wittman, D.~M. et al.\ 2002, SPIE, 4836, 73 
\bibitem{wittman03} Wittman, D. et al.\ 2003, ApJ, 597, 218 
\bibitem{wittman06} Wittman, D. et al.\ 2006, ApJ, 643, 128 
\bibitem{zitrin09} Zitrin, A. et al.\ 2009, MNRAS, 396, 1985 

%
\end{thebibliography}
\end{document}